\documentclass[twocolumn,showpacs,preprintnumbers,amsmath,amssymb]{revtex4}

\usepackage{graphicx}
\usepackage{dcolumn}
\usepackage{bm}

\begin{document}


\title{Optical rotation of heavy hole spins by non-Abelian geometrical means}

\author{Hui Sun$^{1,2}$}
\author{Xun-Li Feng$^{1,2}$}
\author{Chunfeng Wu$^{1,2}$}
\author{Jinming Liu$^{4,1}$}
\author{Shangqing Gong$^{3}$}
\author{C. H. Oh$^{1,2}$}

\affiliation{$^{1}$ Centre for Quantum Technologies, National
University of Singapore, 3 Science Drive 2, Singapore 117543\\
$^{2}$ Department of Physics, National University of
Singapore, 2 Science Drive 3, Singapore 117542\\
$^{3}$ State Key Laboratory of High Field Laser Physics, Shanghai
Institute of Optics and Fine Mechanics, Chinese Academy of Sciences,
Shanghai 201800, China\\
$^{4}$ Physics Department, East China Normal University, Shanghai,
China}

\date{\today}

\begin{abstract}
A non-Abelian geometric method is proposed for rotating of heavy
hole spins in a singly positive charged quantum dot in Voigt
geometry. The key ingredient is the delay-dependent non-Abelian
geometric phase, which is produced by the nonadiabatic transition
between the two degenerate dark states. We demonstrate, by
controlling the pump, the Stokes and the driving fields, that the
rotations about $y$- and $z$-axes with arbitrary angles can be
realized with high fidelity. Fast initialization and heavy hole spin
state readout are also possible.
\end{abstract}

\pacs{78.67.Hc, 03.67.Lx, 03.65.Vf}
\maketitle

\section{introduction}

Electron spins in quantum dots (QDs) are promising candidate for
implementations of
qubits~\cite{loss-pra-1998,imamoglu-prl-1999,hason-rmp-2007} because
of their potential integration into microtechnology. The two spin
states of electron can be mapped directly to the two operational
states in quantum information processing (QIP). A key element for
spin-based QIP is the coherent manipulation of the spin
states~\cite{press-nature-2008,carter-prb-2007,dutt-prb-2006,
coish-prb-2007,hason-prl-2007,wu-prl-2007,
greilich-np-2009,emary-jp-2007,economou-prl-2007,nowack-science-2007,
petta-science-2005,de-prl-2009,andlauer-prb-2009,paspalakis-prb-2004,galland-prl-2008}.
This QIP approach requires not only rotation of unknown spin
states--the heart of spin-based QIP, but also the spin states
initialized in a known state and readout of spins. There has been
significant experimental progress in the demonstration of the key
DiVincenzo requirements~\cite{divincenzo-fp-2000}, for examples,
efficient optical methods for initialization and readout of
spins~\cite{atature-science-2006,emary-prl-2007,atature-np-2007,xu-prl-2007,kim-prl-2008}.
Significant theoretical and experimental effort has been invested in
optical manipulation of electron spin such as by using two
Raman-detuned laser pulses~\cite{emary-jp-2007}, Abelian geometric
phase induced by $2\pi$ pulses~\cite{economou-prl-2007}, resonant
radio-frequency pulses~\cite{nowack-science-2007}, and so on. Using
ultrafast optical pulses, Press {\it et al.} reported that they have
controlled and observed the spin of a single electron in a
semiconductor (over six Rabi oscillations between the two spin
states)~\cite{press-nature-2008}. Subsequently, the rotations of
electron spins about arbitrary axes in a few picoseconds were also
demonstrated in an ensemble of QDs~\cite{greilich-np-2009}.

In spin-based QIP, in addition to preparing the spin in a precisely
defined state, this state should survive long enough to allow its
manipulations. Therefore a long spin coherence time is necessary.
Different from the conduction electron, a valence hole has an atomic
$p$ orbital, which has negligible overlap with the nuclei.
Consequently, the suppressed hyperfine interaction leads to a longer
spin coherence time than that of electron. This may provide an
attractive route to hole-spin-based applications free from the
complications caused by the fluctuating nuclear spin system. In
particular, Heiss {\it et al.} reported that the spin-relaxation
times of holes are up to 270 microseconds in InGaAs QDs embedded in
a GaAs diode structure~\cite{heiss-prb-2007}. Besides long coherence
times, an equally important requirement is the ability to manipulate
spins coherently. More recently, the high-fidelity hole spin
initialization by optical pumping~\cite{gerardot-nature-2008},
optical control and readout of hole spin~\cite{ramsay-prl-2008} have
been demonstrated experimentally. These works promote the spin of a
hole in a semiconductor QD to be the best position to be a contender
for the role of a solid-state qubit.

When a quantum system governed by a Hamiltonian with nondegenerate
eigenstates undergoes some appropriate cyclic evolutions by
adiabatically changing the controllable parameters, besides a
dynamical phase, it may acquire a so-called geometric phase or Berry
phase~\cite{berry-1984}. Wilczek and Zee generalized the geometric
phase to degenerate systems, i.e., non-Abelian geometric
phase~\cite{wilczek-prl-1984}. The geometric phase differs from the
dynamic phase in that the former depends only on the geometry of the
path executed, being therefore insensitive to the local inaccuracies
and fluctuations. They are thus expected to be particularly robust
against
noise~\cite{zanardi-pachos,recati-pra-2002,duan-science-2001,zhu-prl-2003,
feng-pra-2007,solinas-pra-2003}.

Motivated by these work, we propose a method for manipulating
arbitrary rotation of an unknown heavy hole (HH) spin state in a
singly positive charged quantum dot. By applying an external
magnetic field in Voigt geometry, a double tripod-shaped scheme can
be configured. Most importantly, in contrast to the existing
proposals based on electron spin, the HH spin rotations are realized
in terms of the non-Abelian Berry phase, which is acquired by
controlling the parameters along adiabatic loops, i.e., stimulated
Raman adiabatic passage (STIRAP)~\cite{bergmann-rmp-1998} and
fractional STIRAP~\cite{vitanov-jpb-1999}. The STIRAP process can be
used to transfer populations coherently between quantum states
through ``dark state" which efficiently suppress relaxation. The
geometric phases accumulated during a STIRAP process were previously
investigated for tripod systems~\cite{unanyan-pra-1999} and
double-$\Lambda$ systems~\cite{jin-pra-2004}. In this paper, after
briefly reviewing non-Abelian geometric phase
(Sec.~\ref{non-abelian-gp}), we discuss the hole and electron energy
levels of a singly positive charged QD in Voigt geometry and the
selection rules in Sec.~\ref{sec-structre}, and study the
feasibility of initialization by optical pumping. In
Sec.~\ref{sec-rotations} we show how to achieve a two-fold
degenerate dark states, and how to implement the rotations about
$y$- and $z$-axes by using the non-Abelian geometric phase produced
by the nonadiabatically coupling between the two degenerate dark
states. The fidelities of these rotations are also discussed in this
section. The readout of spin state is discussed in
Sec.~\ref{sec-readout}. In Sec.~\ref{sec-summary}, we end with some
remarks.

\section{non-abelian geometric phase}\label{non-abelian-gp}

Our propositions of rotating the HH spin about $y$- and $z$-axes are
based on non-Abelian geometric phase~\cite{wilczek-prl-1984}, so we
start by recalling the basic facts about non-Abelian geometric
phase~\cite{zanardi-pachos}. We consider an $n$-fold degenerate
eigenspace of a Hamiltonian $H(\chi_{\kappa})$
($\kappa=1,2,\cdots,N$) (i.e., the eigenspace information encoded)
depending continuously on parameters $\chi_{\kappa}$. Based on the
time-dependent Schr\"{o}dinger equation, we control the parameters
along loops ${\cal O}$ in an adiabatic fashion so that the initial
preparation can evolve according to
\begin{eqnarray}
|\Psi(t)\rangle=U({\cal O})_{A}|\Psi(t=0)\rangle.
\end{eqnarray}
The transformation can be computed in terms of the Wilczek-Zee gauge
connection~\cite{wilczek-prl-1984}:
\begin{eqnarray}
U({\cal O})_{A}={\cal P}\exp\oint_{\cal
O}\sum\limits_{\kappa=1}^{N}A_{\kappa}d\chi_{\kappa}.\label{evolve-operator}
\end{eqnarray}
${\cal P}$ denotes the path-order operator. $A_{\kappa}^{ab}$ is
called the gauge potential given by
\begin{eqnarray}
A_{\kappa}^{ab}=\left\langle\psi^{a}(\chi)\left|\frac{\partial}{\partial\chi_{\kappa}}\right|\psi^{b}(\chi)\right\rangle,\label{gauge-potential}
\end{eqnarray}
with $\{|\psi^{a}(\chi)\rangle\}_{a=1}^{n}$ being an orthonormal
basis of the degenerate eigenspace. It is worth noting that the
parameter-dependent Hamiltonian should evolve adiabatically so that
the instantaneous state $|\Psi(t)\rangle$ does not overflow the
state vector space spanned by $|\psi^{a}\rangle$. An intriguing
feature of the gauge potential $A$ lies in that it depends only on
the geometry of executed path in the space of degenerate states.

\section{the energy levels and
initialization}\label{sec-structre}

\begin{figure}
\includegraphics[width=8cm]{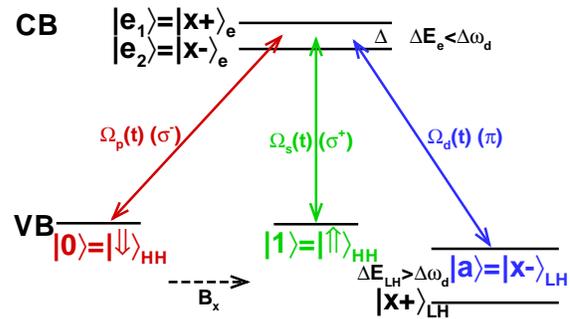}
\caption{(Color online) An energy-level diagram of a singly positive
charged quantum dot in Voigt geometry together with optical spin
selection rules: a degenerate heavy hole (HH) spin states and two
eigenstates of nondegenerate electron and light hole (LH) spin
states. Solid arrows, respectively, indicate dipole allowed optical
transitions driven by the pump ($\Omega_{p}(t)$), the Stokes
($\Omega_{s}(t)$),the driving ($\Omega_{d}(t)$) fields with
$\sigma^{+}$, $\sigma^{-}$ and $\pi$ denoting right-hand, left-hand
circular, and linear-polarized fields. The photon bandwidth
($\Delta\omega_{d}$) should be larger than the electron Zeeman
splitting $\Delta E_{\rm e}$ but smaller than the LH Zeeman
splitting ($\Delta E_{\rm LH}$).}\label{level}
\end{figure}

In the scheme of rotating HH spin, the basic idea is to reconfigure
a multi-level system with interacting Hamiltonian possessing
two-fold degenerate dark states. By changing the Rabi frequencies in
adiabatic fashion, we perform a loop in the parameter space. At both
the beginning and the end of the cycle, we have only the HH spin
states (up and down). But after a loop a non-Abelian geometric phase
is accumulated at HH spin states. Based upon this property,
arbitrary rotations of HH spin can be implemented.

We consider a singly positive charged GaAs/AlGaAs QD with growth
direction $z$, which can be formed naturally by interface steps in
narrow quantum well (QW). The electron and holes become localized
into QD regions of the QW. In the absence of magnetic field, the
lowest conduction-band (CB) level is two-fold degenerate with
respect to the spin projection $\pm1/2$. In the valence-band (VB),
the hole has a total angular momentum of 3/2, with the projection
$m_{J}=\pm1/2$ (``light hole" LH) doublet separated by more than
30-50~meV from the $m_{J}=\pm3/2$ (HH) states due to confinement. It
is very large compared to the bandwidth of the picosecond and
femtosecond pulsed laser, so one should be able to separate the HH
and LH excitations by picosecond pulsed fields in practical
applications. The spin states of HH (spin up and down) trapped in
the QD, which are denoted by
$|0\rangle=|\Downarrow\rangle=|\frac{3}{2},\frac{3}{2}\rangle$ and
$|1\rangle=|\Uparrow\rangle=|\frac{3}{2},-\frac{3}{2}\rangle$, are
our qubit degrees of freedom. We will perform sequentially the
optical initialization, rotations of this spin by non-Abelian
geometrical means, and readout of a single hole spin. With
$\sigma^{-}$ and $\sigma^{+}$ excitations, the only dipole allowed
optical transitions from the valence HH states to the conduction
electron states are
$|\frac{3}{2},\frac{3}{2}\rangle\to|\frac{1}{2},\frac{1}{2}\rangle$
and
$|\frac{3}{2},-\frac{3}{2}\rangle\to|\frac{1}{2},-\frac{1}{2}\rangle$,
and the light hole states cannot be excited because of the frequency
selection. The angular moment restriction inhibits optical coupling
between the two HH spin states. Consequently, the structure can be
described by two degenerate independent two-level systems, and the
spin flip Raman scattering transitions are ideally
dark~\cite{xu-prl-2007}. The dark transitions should become bright
so as to implement spin state initialization, rotations, and
readout. The problem can be solved by applying a magnetic field in
the Voigt geometry ($x$ direction). The magnetic field lifts the
Kramer degeneracy of the LH and electron and reconfigures the
eigenstates as $|x\pm\rangle_{\rm
LH}=(|\frac{3}{2},\frac{1}{2}\rangle\pm|\frac{3}{2},-\frac{1}{2}\rangle)/\sqrt{2}$,
and $|x\pm\rangle_{\rm
e}=(|\frac{1}{2},\frac{1}{2}\rangle\pm|\frac{1}{2},-\frac{1}{2}\rangle)/\sqrt{2}$
(parallel or antiparallel to the magnetic field direction $x$) while
keeping the degeneracy of the HH with a negligibly small in-plane
$g$ factor~\cite{wu-prl-2007,kosaka-prl-2008}. The HH spin states
remain unaffected by the magnetic field. A double tripod-shaped
system is therefore configured, and the energy level diagram and the
associated selection rules are represented in Fig.~\ref{level}. In
the rotations of the spin state, the state $|x-\rangle_{\rm LH}$ is
used as an ancillary level, and it is presented by $|a\rangle$. The
transitions from the VB states $|0\rangle$, $|1\rangle$, and
$|a\rangle$ to the CB states $|e_{1,2}\rangle$ can be excited by the
pump ($\sigma^{-}$-polarized), the Stokes ($\sigma^{+}$-polarized),
and the driving ($\pi$-polarized) fields,
respectively~\cite{book-select-rule}. The dark transitions therefore
become bright and it is possible to initialize and rotate the HH
spin states. The photon bandwidth of the driving field
($\Delta\omega_{d}$) can be controlled in a way such that it is
larger than the electron Zeeman splitting $\Delta E_{\rm e}$ but
smaller than the LH Zeeman splitting ($\Delta E_{\rm
LH}$)~\cite{kosaka-prl-2008}, the excitation of state
$|x+\rangle_{\rm LH}$ can be thus ignored safely.

\begin{figure}
\includegraphics[width=8cm]{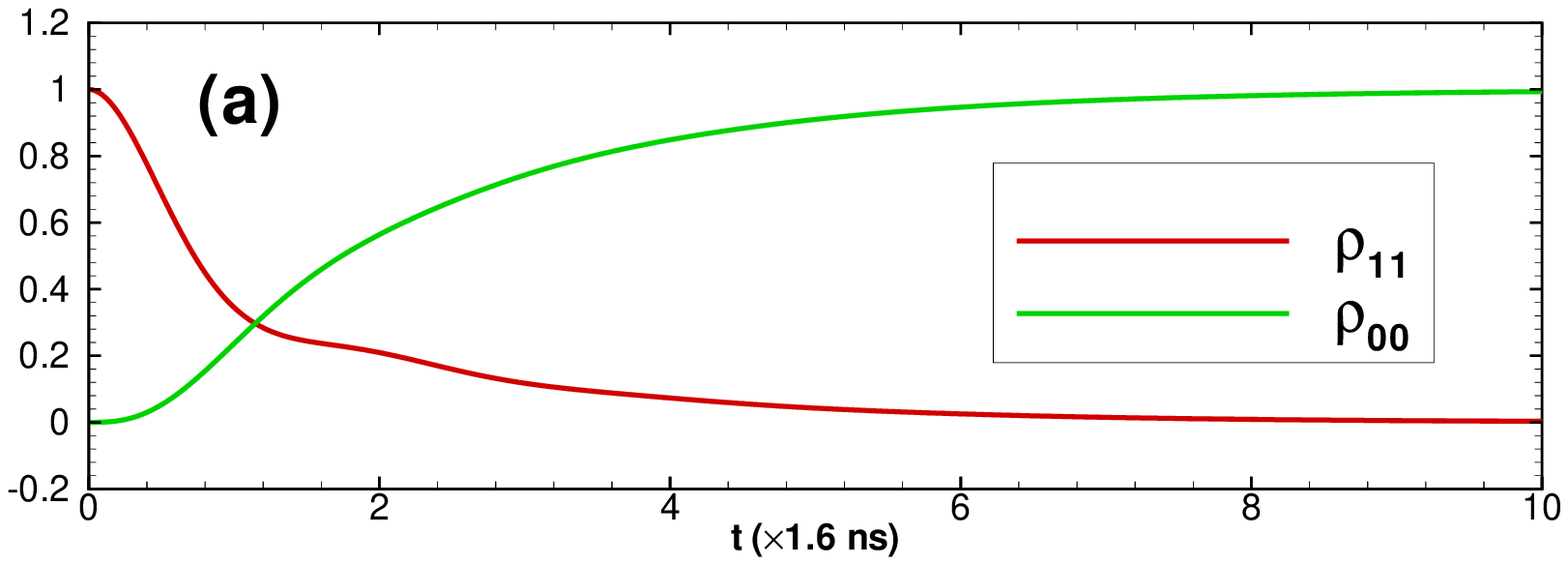}
\includegraphics[width=8cm]{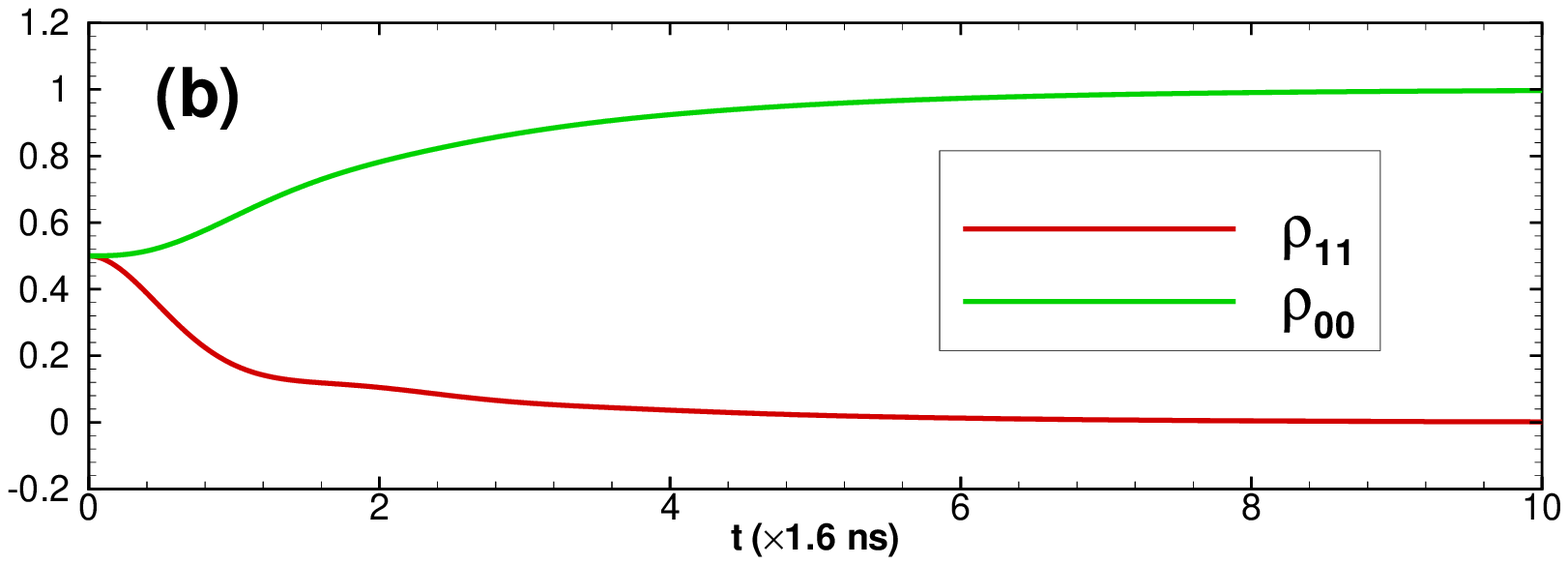}
\caption{(Color online) The population of the states $|0\rangle$ and
$|1\rangle$ as a function of time with $\sigma^{-}$ continuous
illumination. Different initial population distributions are
considered: (a) $\rho_{00}(t=0)=0$, $\rho_{11}(t=0)=1.0$; (b)
$\rho_{00}(t=0)=\rho_{11}(t=0)=0.5$. The values of parameters are
explained in the text.}\label{inilizationalize}
\end{figure}

The initialization of HH spin can be accomplished by optical
pumping~\cite{galland-prl-2008,gerardot-nature-2008}. For example,
the spin up HH state can be prepared by only applying the Stokes
field. The electron is excited from the VB band state to two CB band
states $|e_{1}\rangle$ or $|e_{2}\rangle$, and one spin up HH is
left. Exciton (electron-hole pair) recombination occurs between the
CB band electron and the VB HH with spin down or up, while the
electron recombined with $|\Uparrow\rangle$ hole will be excited
again. Since the pump field has not been applied, the resident hole
remains in spin up state $|\Uparrow\rangle$. Conversely, the HH can
also be prepared in spin down state $|\Downarrow\rangle$ by only
applying the pump field. To assess the efficiency of optical spin
preparation, we have performed numerical simulations using the
Liouville equation for the density matrix for the four-level system
as shown in Fig.~\ref{inilizationalize}. We have assumed that four
exciton recombining channels (from two CB electron states to two
spin states) proceed incoherently, and they are equal to each other
($1/2\gamma=800$~ps~\cite{gerardot-nature-2008}). The magnetic field
and the spin-flip rates for hole and electron are taken as
$B_{x}=55$~mT,
$\gamma_{hh}^{-1}=\gamma_{ee}^{-1}=1$~ms~\cite{gerardot-nature-2008},
and the Rabi frequencies as $1.0\gamma$. The HH spin state
initialization with a fidelity close to 1 [$\sim99.95\%$, we define
the fidelity of the hole spin initialization as
$(\rho_{00}-\rho_{11})/(\rho_{00}+\rho_{11})$ for $\sigma^{+}$
polarization with $\rho_{00}$ ($\rho_{11}$) being the population of
state $|0\rangle$ ($|1\rangle$)] is possible to be achieved within a
few times the inverse of the exciton recombining rate ($\sim1.6$~ns)
in the QD structure in Voigt geometry. The initialization of HH spin
here is realized based on the fact that the magnetic field applied
in Voigt geometry reconfigures the electron eigenstates, which is
different from Refs.~\cite{galland-prl-2008,gerardot-nature-2008},
where the initializations are realized based on spin precession.

\section{arbitrary rotations by non-Abelian geometrical
means}\label{sec-rotations}

It is well known that, with two noncommutable rotations about two
axes, any rotation can be implemented as a composite
rotation~\cite{barenco-pra-1995}. Here we design two noncommutable
rotations about $y$- and $z$-axes and compose general rotations from
them. In the following, based on nono-Alelian geometric phase, we
first show how to rotate the HH spin about $y$-axis, and then
explain how to control the rotation about $z$-axis with the relative
phase between the Stokes and the driving fields. As a result, any
rotations of the HH spin can be realized.

\subsection{Rotation about $y$-axis}

In order to rotate the HH spin about $y$-axis, we apply the pump
field, the Stokes field and the driving field to excite the
corresponding transitions (as shown in Fig.~\ref{level}). The
Hamiltonian in the interaction picture and in the rotating-wave
approximation (RWA) is given by
\begin{eqnarray}
&&\hspace{-0.5cm}H(t)=\hbar[(\Delta_{s}-\Delta_{p})|1\rangle\langle1|+(\Delta_{d}-\Delta_{p})|a\rangle\langle
a|\nonumber\\
&&\hspace{0.7cm}-\Delta_{p}|e_{1}\rangle\langle
e_{1}|-(\Delta_{p}+\Delta)|e_{2}\rangle\langle
e_{2}|-\Omega_{p1}(t)|e_{1}\rangle\langle0|\nonumber\\
&&\hspace{0.7cm}-\Omega_{s1}(t)|e_{1}\rangle\langle1|-\Omega_{d1}(t)|e_{1}\rangle\langle a|-\Omega_{p2}(t)|e_{2}\rangle\langle0|\nonumber\\
&&\hspace{0.7cm}-\Omega_{s2}(t)|e_{2}\rangle\langle1|-\Omega_{d2}(t)|e_{2}\rangle\langle
a|]+{\rm H.c.},\label{hamiltonian-y}
\end{eqnarray}
where the half Rabi frequency is defined as $\Omega_{jk}(t)=\langle
e_{k}|\vec{\mu}\cdot\vec{E}_{j}(t)|q\rangle/2\hbar$ with $\vec{\mu}$
being the dipole moment, $k=1,2$, $j=p(s,d)$ denoting the pump (the
Stokes and the driving) field, and $q=0,1,a$.
$\Delta_{j}=\omega_{j}-(\omega_{e1}-\omega_{q})$ is the detuning and
$\Delta=\omega_{e_{1}}-\omega_{e_{2}}=|g_{x}^{e}|\mu_{B}B_{x}/\hbar$
is the electron Zeeman splitting with $g_{x}^{e}$ and $\mu_{B}$
representing Land\'{e} factor of electron and Bohr magneton,
respectively.

When the pump, the Stokes, and the driving fields are tuned to match
the conditions $\Delta_{p}=\Delta_{s}=\Delta_{d}\neq-\Delta/2$
(three fields are tuned to three-photon resonance, but they are not
at the middle point of the two electron Zeeman splitting levels) and
$\Omega_{j1}(t)/\Omega_{j2}(t)=C$ (for simplicity, we choose $C=1$
and denote $\Omega_{j1}(t)=\Omega_{j2}(t)=\Omega_{j}(t)$), one can
easily find from the interaction Hamiltonian (\ref{hamiltonian-y})
that the interacting system has two degenerate dark states
\begin{eqnarray}
&&|D_{1}\rangle=\cos\theta(t)|1\rangle-\sin\theta(t)|a\rangle,\label{dark-1-y}\\
&&|D_{2}\rangle=\cos\varphi(t)|0\rangle-\sin\varphi(t)\sin\theta(t)|1\rangle\nonumber\\
&&\hspace{1.2cm}-\sin\varphi(t)\cos\theta(t)|a\rangle,\label{dark-2-y}
\end{eqnarray}
where the mixing angle $\theta(t)$ and the additional mixing angle
$\varphi(t)$ are defined as
\begin{eqnarray}
\tan\theta(t)=\frac{\Omega_{s}(t)}{\Omega_{d}(t)},\hspace{0.4cm}\tan\varphi(t)=\frac{\Omega_{p}(t)}{\sqrt{\Omega_{s}^{2}(t)+\Omega_{d}^{2}(t)}}.
\end{eqnarray}
It is well know that the exciton recombination occurs only if there
is electron excited to the CB states $|e_{1,2}\rangle$. The two-fold
degenerate dark states $|D_{1,2}\rangle$, which are known as trapped
states, receive no contributions from the CB electron states [see
Eqs.~(\ref{dark-1-y}) and (\ref{dark-2-y})]. Hence the rotation of
HH spin about $y$-axis is robust against the exciton recombination
process, and thus leading to high fidelity operations. It is also
worth noting that, in the absence of the three fields, i.e., all the
parameters (actually the angles $\theta(t)$ and $\varphi(t)$) are
fixed to zero, the previous eigenstates coincide with the two spin
states $|D_{1}(0)\rangle=|\Uparrow\rangle$ and
$|D_{2}(0)\rangle=|\Downarrow\rangle$. When the three fields are
applied adiabatically and hence the angles $\theta(t)$ and
$\varphi(t)$ change adiabatically, the non-Abelian geometric
connection components can be calculated, according to
Eq.~(\ref{gauge-potential}), and we have
\begin{eqnarray}
A=A_{\theta}d\theta=-i\sin\varphi(t)\sigma_{y}d\theta,\label{gauge-potential-ry}
\end{eqnarray}
with $\sigma_{y}$ being the $y$-component Pauli matrix. The related
unitary operation is
\begin{eqnarray}
U({\cal O})=\exp\left(-i\sigma_{y}\int_{{\cal O}}\sin\varphi(t)
d\theta\right)=R_{y}(\beta),\label{eq-rotation-y}
\end{eqnarray}
where the rotating angle $\beta$ is given by
\begin{eqnarray}
\beta=\int_{\cal O}\sin\varphi(t) d\theta.
\end{eqnarray}
Similar degenerate dark states and non-Alelian geometric connection
have been realized for ion trap~\cite{duan-science-2001} and
atoms~\cite{zanardi-pachos,unanyan-pra-1999}. However, in the QD
structure under consideration, the non-Abelian geometric connection
$A$ is based on HH spin states.

\begin{figure}
\includegraphics[width=8cm]{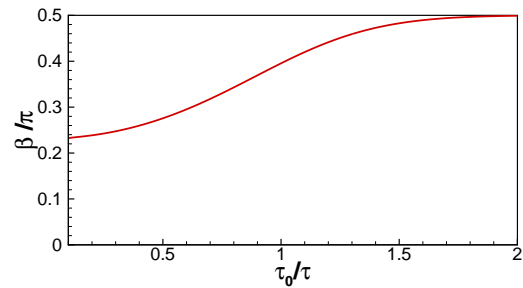}
\caption{(Color online) The rotating angle $\beta$ about $y$-axis in
units of $\pi$ as a function of the delay $\tau_{0}$. The values of
parameters are explained in the text.}\label{fig-gp-ry}
\end{figure}

For a quantitative analysis of the rotating angle $\beta$ about
$y$-axis, we assume that the pump, the Stokes and the driving fields
have Gaussian shapes as
$\Omega_{d}(t)=\Omega_{d}^{0}\exp[-(t+\tau_{0})^{2}/\tau^{2}]$,
$\Omega_{p}(t)=\Omega_{p}^{0}\exp(-t^{2}/\tau^{2})$, and
$\Omega_{s}(t)=\Omega_{s}^{0}\exp[-(t-\tau_{0})^{2}/\tau^{2}]$ with
$\tau$ and $\tau_{0}$ being, respectively, the pulse widths and the
delay. Figure~\ref{fig-gp-ry} shows that the evolution of the
rotating angle $\beta$ about $y$-axis in unit of $\pi$ as a function
of delay $\tau_{0}$. In the interaction, the time-dependent
Hamiltonian should be performed sufficiently slowly. According to
the condition for adiabatic passage~\cite{bergmann-rmp-1998}, we
take
$\Omega_{p}^{0}\tau=\Omega_{s}^{0}\tau=\Omega_{d}^{0}\tau=50\gg1$,
which ensures no transition between the dark and bright states (not
given in this paper). Figure~\ref{fig-gp-ry} shows clearly that the
rotating angle $\beta$ is delay-dependent. It goes up successively
with increasing value of the delay, and then reaches its maximum
value, $\pi/2$, when the delay $\tau_{0}$ is large.

Thus we have shown that the rotation of hole spin about $y$-axis can
be implemented by using the non-Abelian geometric phase. The
rotation is determined only by the global property and does not
depend upon the details of the evolution path in the parameters
space.

\subsection{Rotation about $z$-axis}

As suggested in Ref.~\cite{duan-science-2001}, by setting
$\Omega_{p}(t)=0$ and changing adiabatically $\Omega_{s}(t)$ and
$\Omega_{d}(t)$, the rotation about $z$-axis can be achieved by
making use of the Abelian geometric phase in our QD system. However,
by involving the non-Abelian geometric phase, here we suggest
another method for rotating the hole spin about $z$-axis. An
important feature of this rotation is that the rotating angle about
$z$-axis is not geometric phase-dependent, it is controlled by the
relative phase between the Stokes and the driving fields. To do so,
we set $\Omega_{p}(t)=0$ so that the spin down state $|0\rangle$ is
decoupled, and assume the phase of the driving field is zero, the
relative phase $\phi$ is therefore the phase of the Stokes field.
The time-dependent Hamiltonian $H(t)$ in the interaction picture and
RWA takes the form
\begin{eqnarray}\label{hamiltonian-z}
&&\hspace{-0.4cm}H(t)=\hbar[(\Delta_{d}-\Delta_{s})|a\rangle\langle
a|-\Delta_{s}|e_{1}\rangle\langle
e_{1}|\nonumber\\
&&\hspace{0.8cm}-(\Delta_{s}+\Delta)|e_{2}\rangle\langle
e_{2}|-\Omega_{d}(t)(|e_{1}\rangle\langle a|+|e_{2}\rangle\langle
a|)\nonumber\\
&&\hspace{0.8cm}-\Omega_{s}(t)\exp(-i\phi)(|e_{1}\rangle\langle
1|+|e_{2}\rangle\langle 1|)]+{\rm H.c.},
\end{eqnarray}
where $\phi$ is the relative phase between the Stokes and the drive
fields. In the derivation of Hamiltonian (\ref{hamiltonian-z}), the
condition $\Omega_{j1}/\Omega_{j2}=C=1$ is applied. When the Stokes
and the driving fields are controlled to satisfy two photon
resonance, only one dark state ($|D_{1}(t)\rangle$) exists. The hole
spin cannot be rotated by non-Abelian geometrical means.
Fortunately, however, when the two fields are tuned to the middle
point of the two electron Zeeman splitting levels, there is another
dark state~\cite{jin-pra-2004}. The two-fold degenerate dark states
are
\begin{eqnarray}
&&\hspace{-0.5cm}|D_{1}(t)\rangle=\cos\theta(t)e^{i\phi}|1\rangle-\sin\theta(t)|a\rangle,\label{dark-1-z}\\
&&\hspace{-0.5cm}|D_{2}(t)\rangle=\frac{1}{\sqrt{2}}\cos\varphi(t)(|e_{1}\rangle-|e_{2}\rangle)+\sin\varphi(t)\cos\theta(t)|a\rangle\nonumber\\
&&\hspace{1.2cm}+\sin\varphi(t)\sin\theta(t)e^{i\phi}|1\rangle,\label{dark-2-z}
\end{eqnarray}
where the mixing angle $\theta(t)$ and the additional mixing angle
$\varphi(t)$ related to the electron Zeeman energy splitting
$\Delta$ are respectively defined as
\begin{eqnarray}
\tan\theta(t)=\frac{\Omega_{s}(t)}{\Omega_{d}(t)},\hspace{0.4cm}
\tan\varphi(t)=\frac{\Delta/2}{\sqrt{2(\Omega_{s}^{2}(t)+\Omega_{d}^{2}(t))}}.
\end{eqnarray}
A similar form of double degenerate dark states in double-$\Lambda$
atomic scheme has been obtained~\cite{jin-pra-2004}. However, in our
QD structure the degenerate HH spin-based dark states stem from
electron Zeeman splitting.

In adiabatic limit (the time derivative of the mixing angles
$\theta(t)$ and $\varphi(t)$ are small compared with the splitting
of eigenvalues, given by
$2\sqrt{2(\Omega_{p}^{2}(t)+\Omega_{s}^{2}(t))+(\Delta/2)^{2}}$),
only the transitions between the degenerate dark states
$|D_{1}(t)\rangle$ and $|D_{2}(t)\rangle$ should be taken into
account, and the nonadiabatic coupling of states $|D_{1}(t)\rangle$
and $|D_{2}(t)\rangle$ to other states (the expressions have not
given in this paper) can be safely ignored~\cite{unanyan-pra-1999}.
$\langle D_{2}|\dot{D}_{1}\rangle=-\sin\varphi(t)\dot{\theta}(t)$
also exhibits that a nonadidabatic transition between the two-fold
degenerate dark states may occur. Although the dark state
$|D_{2}(t)\rangle$ receives contribution from the CB electron states
[see Eq.~(\ref{dark-2-z})], we note that the electron
magnetic-dependent Zeeman energy splitting is independent of time.
At the beginning and end of the adiabatic process, we have
$\Delta^{2}\gg\Omega_{s}^{2}(t)+\Omega_{d}^{2}(t)$, which therefore
leads to $|\varphi|=\pi/2$. The CB electron states have no influence
on the final dark state $|D_{2}(+\infty)\rangle$. In other words,
there is no electron in the CB states when the interaction is
finished.

\begin{figure}
\includegraphics[width=8cm]{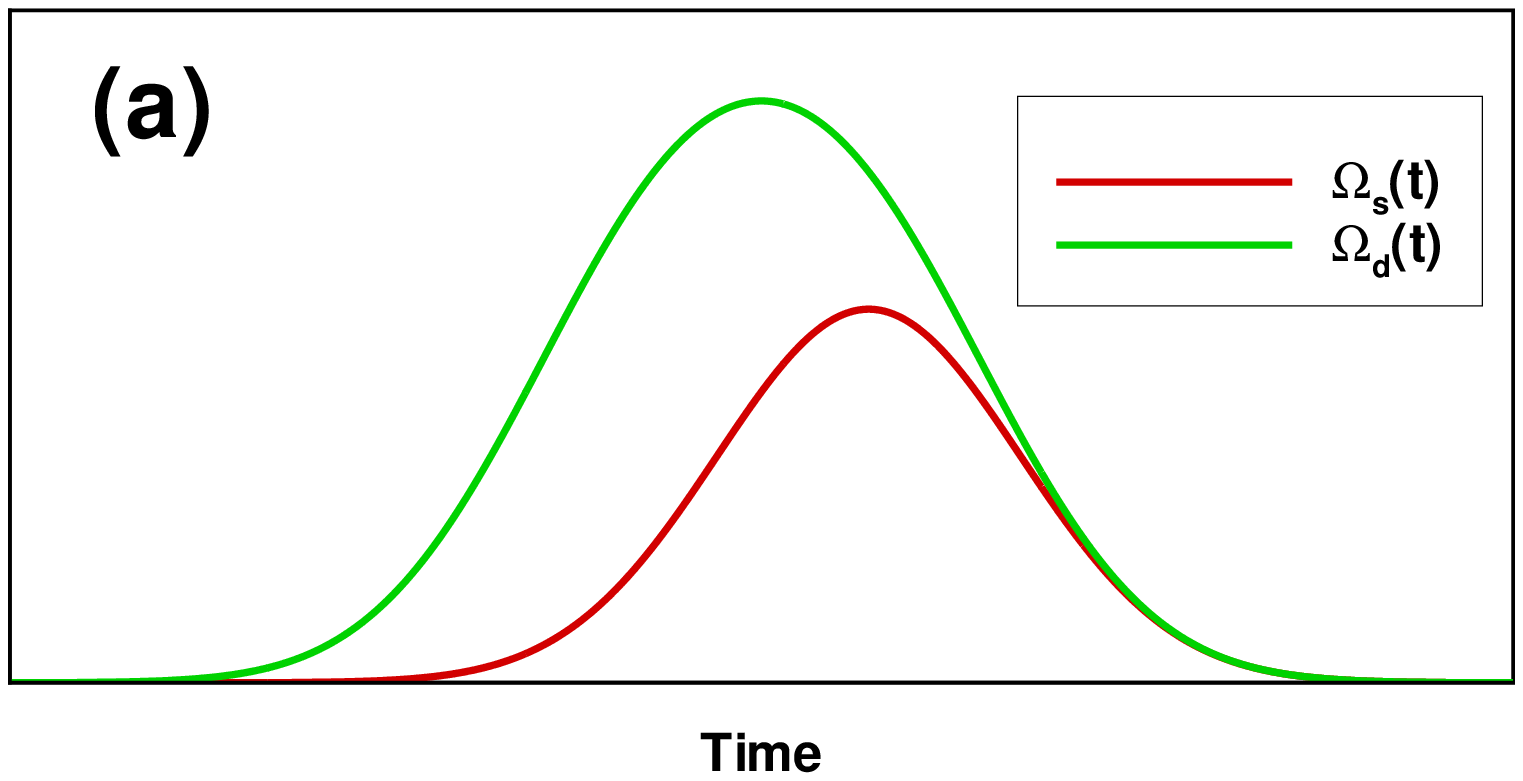}
\includegraphics[width=8cm]{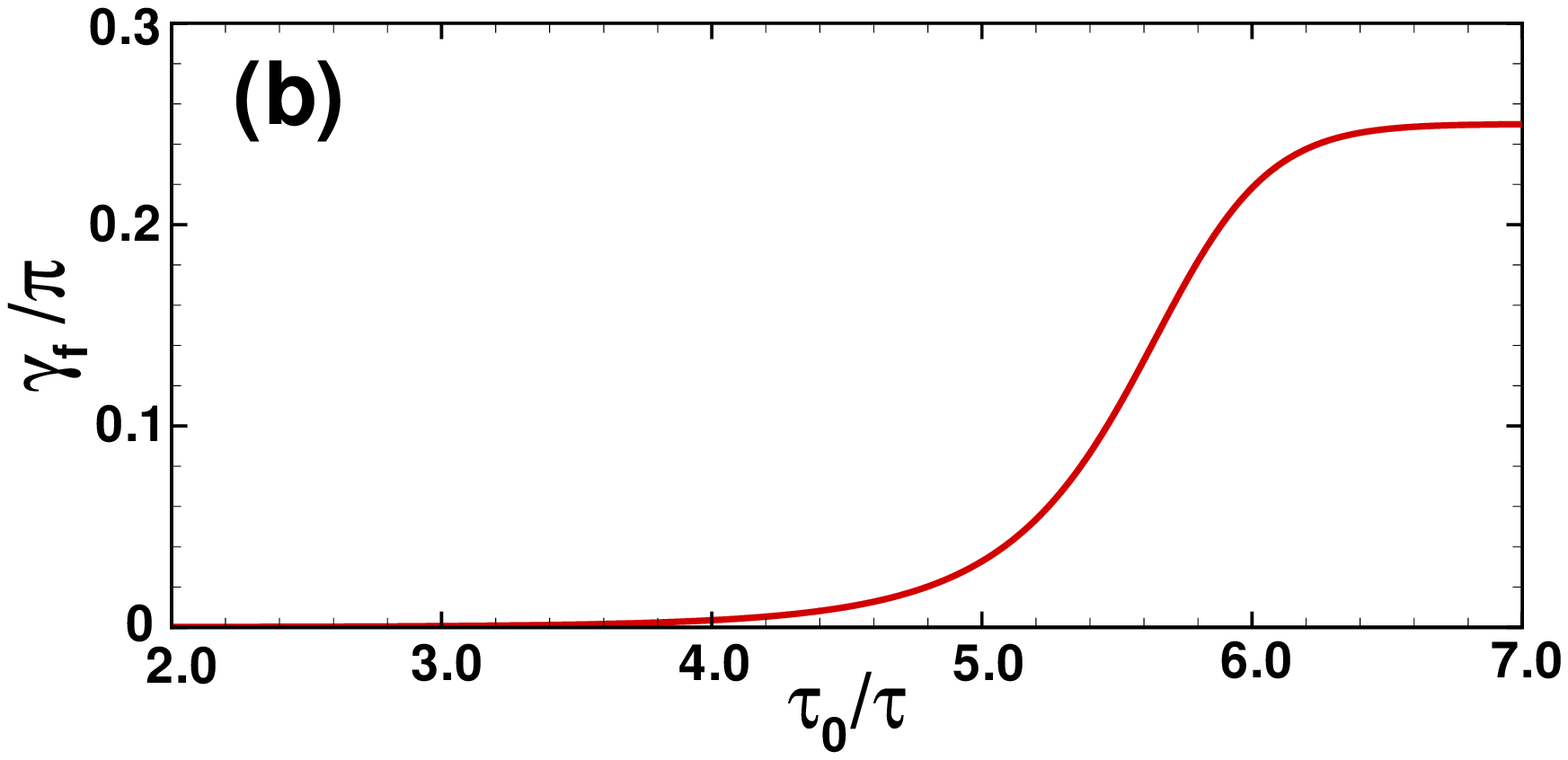}
\caption{(Color online) (a) The shapes of the Stokes and the driving
fields defined by Eqs.~(\ref{omegas}) and (\ref{omegad}). (b) The
geometric phase $\gamma_{f}$ in units of $\pi$ as function of the
delay between two pulses $\tau_{0}$ with
$\Omega_{p}^{0}\tau=\Omega_{s}^{0}\tau=50$. The values of parameters
are explained in the text.}\label{fig-gp-rz}
\end{figure}

Next, we will show how to translate the phase $\phi$ into the HH
spin state $|\Uparrow\rangle$, thus leading to the rotation about
$z$-axis with angle $\phi$. Recalling the fact that, in the
subspace, HH is prepared with spin up, namely
$|D_{1}(0)\rangle=|1\rangle$. The Stokes and the driving fields are
applied in the counterintuitive order, while they terminate with a
constant ratio of their amplitude so that the phase $\phi$ can be
introduced into the HH spin state. This extension of STIRAP is
called fractional STIRAP, which has been suggested to create the
coherent atomic superpositions in a robust
way~\cite{vitanov-jpb-1999}. As shown in Fig.~\ref{fig-gp-rz}(a),
the driving field consists of two parts$-$one with the same time
dependence as the Stokes field and the other coming earlier, for
example
\begin{eqnarray}
&&\hspace{-0.6cm}\Omega_{s}(t)=\Omega_{s}^{0}\exp(-t^{2}/\tau^{2}),\label{omegas}\\
&&\hspace{-0.6cm}\Omega_{d}(t)=\Omega_{d}^{0}\{\exp[-(t+\tau_{0})^{2}/\tau^{2}]+\exp(-t^{2}/\tau^{2})\}.\label{omegad}
\end{eqnarray}
Here $\Omega_{s}^{0}$ and $\Omega_{d}^{0}$ are amplitudes of the
Stokes and the driving fields, $\tau$ and $\tau_{0}$ are,
respectively, pulse widths and delay between the two parts of
driving field. During the interaction, the mixing angle $\theta(t)$
varies from 0 to $\pi/4$. According to the theory of non-Abelian
geometric phase, after the interaction, we
have~\cite{unanyan-pra-1999,jin-pra-2004}
\begin{eqnarray}
&&\hspace{-1cm}|\Psi(+\infty)\rangle=\frac{1}{\sqrt{2}}\left[e^{i\phi}(\sin\gamma_{f}+\cos\gamma_{f})|1\rangle\right.\nonumber\\
&&\left.\hspace{1.1cm}+(\sin\gamma_{f}-\cos\gamma_{f})|a\rangle\right],\label{eq-rotation-d1}
\end{eqnarray}
where $\gamma_{f}$ is given by
\begin{eqnarray}
\hspace{-0.3cm}\gamma_{f}=\oint_{\cal
O}\frac{\Delta/2}{\Omega_{p}^{2}(t)+\Omega_{s}^{2}(t)}\frac{\Omega_{s}(t)d\Omega_{p}(t)-\Omega_{p}(t)d\Omega_{s}(t)}
{\sqrt{2(\Omega_{p}^{2}(t)+\Omega_{s}^{2}(t))+(\Delta/2)^{2}}}.\label{gamma-f}
\end{eqnarray}
Equation~(\ref{eq-rotation-d1}) exhibits that the relative phase
$\phi$ translates to the HH spin state $|\Uparrow\rangle$ via
fractional STIRAP. If $\gamma_{f}$ can be accumulated to $\pi/4$ by
controlling the Stokes and drive fields, after the cycle evolution,
the HH spin will return to the spin up state with the phase $\phi$.
As a result, the rotation of HH spin about $z$-axis is realized, and
the varying rotating angle can be obtained by changing the relative
phase $\phi$, namely the phase of the Stokes field.

It should be noted that $\gamma_{f}$ is gauge invariant, it depends
upon the delay $\tau_{0}$~\cite{unanyan-pra-1999,jin-pra-2004}. In
Fig.~\ref{fig-gp-rz}(b), we plot the evolution of $\gamma_{f}$ in
units of $\pi$ as a function of $\tau_{0}$. We take the Rabi
frequency amplitudes as:
$\Omega_{s}^{0}=\Omega_{d}^{0}=0.5$~ps$^{-1}$, the pulse width
parameter $\tau=100$~ps. The magnetic field is taken as
$B_{x}=55$~mT (we take the in-plane $g$ factor of electron as
$g_{x}^{e}=-0.21$~\cite{kosaka-prl-2008}, then the electron Zeeman
splitting $\Delta\approx1$~GHz). With these parameters, the
adiabatic condition is hold and the governing Hamiltonian evolutes
adiabatically. As shown in Fig.~\ref{fig-gp-rz}(b) $\gamma_{f}$
increases from 0 by degrees with the growing of the delay, and
reaches its maximum value, $\pi/4$, when the delay $\tau_{0}$ is
large. Thus, the relative phase $\phi$ can be used to control the HH
spin rotation about $z$-axis.

By combining the above rotations about $y$- and $z$-axis, any
rotation can be implemented. For example, rotation about $x$-axis
can be realized by
$R_{x}(\phi)=R_{y}^{\dagger}(\pi/2)R_{z}(\phi)R_{y}(\pi/2)$. Our
rotation procedure is sensitive to the non-Abelian geometric phase
and the relative phase between the Stokes and the driving fields.
Moreover, the non-Abelian geometric phase is independent of pulse
areas and dependent on the ratios $\tau_{0}/\tau$,
$\Omega_{s}^{0}/\Delta$, and $\Omega_{d}^{0}/\Delta$. Thus it is
robust against the fluctuation of the pulse shapes, pulse areas and
noise.

\subsection{Fidelity}

\begin{table}
\begin{scriptsize}
\caption{\label{tab-fedility} Fidelity of selected rotations of HH
spin.}
\begin{ruledtabular}
\begin{tabular}{ccc}
$R_{n}(\phi)$&$\tau_{0}/\tau$&Fidelity\\
\cline{1-3} $R_{y}(\pi/2)$ & 1.5 & 99.96\% \\
$R_{z}(\pi/2)$ & 6.5 & 99.99\% \\
$R_{x}(\pi/2)$ &  & 99.94\%\\
\end{tabular}
\end{ruledtabular}
\end{scriptsize}
\end{table}

Based on the non-Abelian geometric phase, arbitrary rotations of HH
spin are possible. But how about the degree to which our approximate
description matches the actual behavior of the system? The fidelity
is a measure of how accurate the target gate is implemented and it
is defined as ${\cal
F}(U)=\overline{|\langle\Psi|U^{\dagger}U_{id}|\Psi\rangle|^{2}}$,
where $U_{id}$ is the target operation, $U$ is the actual operation,
and the average is taken over all input spin
states~\cite{bowdrey-pla-2002}. By numerical simulation of the
density matrix equations, the fidelity can be calculated. Typical
rotation fidelities, listed in Table~\ref{tab-fedility}, are on the
order of $99.9\%$. The rotations of HH spin are realized with high
fidelity.

\section{Readout}\label{sec-readout}

The accurate measurement of the spin state of each qubit is
essential in a quantum computation scheme. In our scheme, the
$\sigma^{+}$-polarized Stokes (the $\sigma^{-}$-polarized pump)
field can only excite the HH state with spin down (up). The
measurement of the HH spin states therefore can be achieved by
applying a $\sigma^{+}$ or $\sigma^{-}$ polarized continuous laser
field. When the $\sigma^{+}$-polarized field is applied, for
example, if the spin is rotated to $|\Downarrow\rangle$, the QD will
emit a single photon from the $|e_{1,2}\rangle\to|\Downarrow\rangle$
transitions, which can be detected using a single-photon
counter~\cite{press-nature-2008}.

\section{conclusions}\label{sec-summary}

To summarize, we consider a singly positive charged quantum dot, and
demonstrate sequentially the initialization, the optical rotations
of HH spin with non-Abelian geometrical means, and readout of a
single hole spin. Together with an magnetic field applied in Voigt
geometry, the quantum dot system can be reconfigured as a double
tripod scheme. When the pump, the Stokes and the driving fields are
tuned to satisfy certain conditions, the QD system has two-fold
degenerate dark states. Based on the non-Abelian geometric phase
produced by the nonadiabatic coupling between the two dark states,
not only can the HH spin be rotated about $y$-axis with stimulated
Raman adiabatic passage, but also the relative phase between the
Stokes and the driving fields can be translated into the hole spin
state with fractional stimulated Raman adiabatic passages, leading
to the implementation of rotation about $z$-axis. Therefore the key
step of optical arbitrary rotations of HH spin with high fidelity
for QIP can be implemented by non-Alelian geometrical means. It is
in principle useful for spin-based quantum information processing.

\begin{acknowledgements}
This work is supported by the National Research Foundation and
Ministry of Education, Singapore under academic research grant No.
WBS: R-710-000-008-271. One of the authors (H. S.) would also like
to acknowledge the support of the National Basic Research Program of
China (973 Program) under Grant No. 2006CB921104, the author (S. Q.
G.) acknowledges funding from the National Natural Science
Foundation of China through Grant No. 10874194 and the author (J. M.
L.) acknowledges funding from the National Natural Science
Foundation of China under Grant No. 60708003.
\end{acknowledgements}

\end{document}